\documentclass[twocolumn,showpacs,preprintnumbers,amsmath,amssymb,superscriptaddress]{revtex4}
\usepackage{graphicx}
\begin{document}
\title{X-ray spectroscopic study of BaFeO$_{3}$ thin films; an Fe$ ^{4+} $ ferromagnetic insulator} 
\author{T. Tsuyama}
\affiliation{Department of Applied Physics and Quantum-Phase Electronics Center (QPEC),\\ University of Tokyo, Tokyo 113-8656, Japan.}
\author{T. Matsuda}
\affiliation{Department of Applied Physics and Quantum-Phase Electronics Center (QPEC),\\ University of Tokyo, Tokyo 113-8656, Japan.}
\author{S. Chakraverty}
\affiliation{RIKEN Center for Emergent Matter Science (CEMS), Wako 351-0198 Japan.}
\author{J. Okamoto}
\affiliation{Condensed Matter Research Center and Photon Factory, Institute of Materials Structure Science (IMSS), High Energy Accelerator Research Organization (KEK), Ibaraki 305-0801, Japan.}
\author{\\E. Ikenaga}
\affiliation{Japan Synchrotron Radiation Research Institute (JASRI)/SPring-8, Kouto 679-5198, Japan.}
\author{A. Tanaka}
\affiliation{Department of Quantum Matter, ADSM, Hiroshima University, Higashi-Hiroshima 739-8530, Japan.}
\author{T. Mizokawa}
\affiliation{Department of Complexity Science and Engineering, University of Tokyo, Chiba 277-8561, Japan.}
\author{H. Y. Hwang}
\affiliation{RIKEN Center for Emergent Matter Science (CEMS), Wako 351-0198 Japan.}
\affiliation{Stanford Institute for Materials and Energy Sciences,\\
SLAC National Accelerator Laboratory, Menlo Park, CA 94025, USA.}
\affiliation{Department of Applied Physics, Stanford University, Stanford, CA 94305, USA.}
\author{Y. Tokura}
\affiliation{Department of Applied Physics and Quantum-Phase Electronics Center (QPEC),\\ University of Tokyo, Tokyo 113-8656, Japan.}
\affiliation{RIKEN Center for Emergent Matter Science (CEMS), Wako 351-0198 Japan.}
\author{H. Wadati}
\affiliation{Department of Applied Physics and Quantum-Phase Electronics Center (QPEC),\\ University of Tokyo, Tokyo 113-8656, Japan.}
\date{\today}
\begin{abstract}
We investigated the electronic and magnetic properties of fully oxidized BaFeO$_{3}$ thin films, which show ferromagnetic-insulating properties with cubic crystal structure, by hard x-ray photoemission  spectroscopy (HAXPES), x-ray absorption spectroscopy (XAS) and soft x-ray magnetic circular dichroism (XMCD). We analyzed the results with configuration-interaction (CI) cluster-model calculations for Fe$ ^{4+} $, which showed good agreement with the experimental results. We also studied SrFeO$ _{3} $ thin films, which have an Fe$ ^{4+} $ ion helical magnetism in cubic crystal structure, but are metallic at all temperatures. We found that BaFeO$ _{3} $ thin films are insulating with large magnetization ($ 2.1 \mu_{B}/\textit{formula\;unit}$) under $ \sim $ 1 T, using valence-band HAXPES and Fe $ 2p $ XMCD, which is consistent with the previously reported resistivity and magnetization measurements. 
Although Fe $ 2p $ core-level HAXPES and Fe $ 2p $ XAS spectra of BaFeO$ _{3} $ and SrFeO$ _{3} $ thin films are quite similar,  we compared the insulating BaFeO$ _{3} $ to metallic SrFeO$ _{3} $ thin films with valence-band HAXPES.
The CI cluster-model analysis indicates that the ground state of BaFeO$ _{3} $ is dominated by $d^{5}\underline{L}$ ($\underline{L}$: ligand hole) configuration due to the negative charge transfer energy, and that the band gap has significant O $ 2p $ character. We revealed that the differences of the electronic and magnetic properties between BaFeO$ _{3} $ and SrFeO$ _{3} $ arise from the differences in their lattice constants, through affecting the strength of  hybridization and bandwidth.
\end{abstract}
\maketitle

%
%
\section{Introduction}
Strongly correlated transition-metal oxides show  interesting physical properties such as colossal magnetoresistance,  metal-insulator transitions, and the ordering of charge, orbital and spin \cite{Imada}. 
Among this class of materials, high-valent metal compounds, such as SrFeO$ _{3} $ and CaFeO$ _{3} $, have attracted great attention due to their rich and anomalous physical properties. \cite{CFO0,CFO00,CFO,CFORaman,SFO,SFOnutron,Bocquet,Abbate,Ishiwata,SuvankarSFO}.
CaFeO$_{3}$ shows charge ordering below 290 K with the charge disproportion of $d^{5}\underline{L}\,+\,d^{5}\underline{L}$ $\longrightarrow$ $d^{5}$ + $d^{5}\underline{L}^{2}$  ($\underline{L}$: ligand hole), inducing monoclinic crystal distortion \cite{CFO00,CFO,CFORaman}.
The ground state of SrFeO$_{3}$ is dominated by the \textit{$d^{5}\underline{L}$} configuration and is metallic with cubic crystal structure. 
The $d^{5}\underline{L}$ electronic configuration gives rise to features in SrFeO$_{3}$, such as the helicoidal magnetic order, metallicity and absence of Jahn-Teller effect, which cannot be expected in the \textit{$d^{4}$} electronic configuration \cite{SFO,Abbate,Bocquet,Mostovoy,ZSA}.
Mostovoy theoretically revealed that the helicoidal magnetic order of perovskite Fe$ ^{4+} $ compound is attributed to the O $ 2p $ hole, rather than the competition between ferromagnetic double-exchange and antiferromagnetic superexchange interaction\cite{Mostovoy}.
Furthermore, the reports of topological Hall effect suggest many possibilities for complex spin textures \cite{Ishiwata,SuvankarSFO}. 

BaFeO$_{3}$ is also one of those Fe$^{4+}$ series, and Hayashi\textit{ et al.} reported the first success in synthesis of polycrystalline cubic BaFeO$_{3}$ with the lattice constant of 3.97 \AA{} \cite{BulkBFO}.
A-type helicoidal magnetic state with $ \textit{\textbf{q}} = (0\;0\;0.06)\,\dfrac{2\pi}{a}$ ($ a $ = 3.97 \AA{}) was found,  but it  becomes ferromagnetic under a small magnetic field of $ \sim $ 0.3 T with a saturation magnetization of 3.5 $\mu_{B}/formula\;unit$ ($f.u.$), and transition temperature of 111 K, respectively \cite{BulkBFO}.
Li $ \textit{et al} $. theoretically showed that the energy difference between the A-type helicoidal magnetic and ferromagnetic phase is quite small, and can be switched by small magnetic field  in BaFeO$ _{3} $ unlike SrFeO$ _{3} $ \cite{TheoreticalBFO,TheoreticalBFO2}.
The difference in their magnetic properties is explained by the difference in their lattice constant: the larger Ba stretches out the lattice and decreases the strength of Fe $ 3d $ - O $2p  $ hybridization, which closely relates to the magnetic phases \cite{Mostovoy,TheoreticalBFO,TheoreticalBFO2}. 

There have been several reports on BaFeO$ _{3} $ thin films on SrTiO$_{3}$ (001) substrates prior to the successful fabrication of bulk BaFeO$ _{3} $, but all of them showed  much lower magnetization and larger lattice constant than bulk BaFeO$_{3}$, due to oxygen vacancies in the thin films \cite{BFO/STO1,BFO/STO2,BFO/STO3,BFO/STO4}.
Chakraverty $\textit{et al.}$ reported the synthesis of fully oxidized single crystalline BaFeO$ _{3} $ thin film which shows large saturation magnetization (in-plane; $3.2\mu_{B}/f.u.$, out-of-plane; $2.7\,\mu_{B}/f.u.$) with the bulk lattice constant of 3.97 \AA{} in cubic crystal structure \cite{Suvankar}. 
Although the saturation magnetization and lattice constant of the recently reported thin film were quite close to that of bulk BaFeO$ _{3} $, no signature of helicoidal magnetic structure has been observed \cite{BulkBFO,Suvankar}. Transport and optical properties indicate that the fully oxidized film is an insulator with band gap $ \sim  $ 1.8 eV.
The absence of the helicoidal magnetic order in BaFeO$_{3}$ film is likely due to the small energy barrier between the A-type helicoidal magnetic and the ferromagnetic phases, so that residual strain, or reduced dimensionality, stabilizes the uniform ferromagnet \cite{TheoreticalBFO2}.

Given these exciting preliminary results on newly available single crystalline BaFeO$ _{3} $ films, spectroscopic studies of the electronic and magnetic structures would be invaluable. 
For this purpose, we carried out hard x-ray photoemission spectroscopy (HAXPES) for Fe $ 2p $ core level and valence band, Fe $2p$ and O $1s$ x-ray absorption spectroscopy (XAS) and x-ray magnetic circular dichroism (XMCD).
The results were analyzed with configuration-interaction (CI) cluster-model calculations.
These experimental and calculated results accurately explain that the difference of the valence-band spectra between BaFeO$_{3}$ and SrFeO$_{3}$ originates from the differences in their strength of Fe $3d  $ - O $ 2p $ hybridization due to the difference in their atomic distances ($R$) [the atomic distance of  BaFeO$ _{3} $ ($ R_{BFO} $) is $ \sim $ 1.99 \AA{} and that of SrFeO$ _{3} $ ($ R_{SFO}$) is  $\sim $ 1.93 \AA{}, respectively].
%
\section{Experimental and calculations}
A single crystalline, cubic, fully oxidized 50 nm thin film of BaFeO$ _{3} $ (with lattice constant 3.97 \AA) was grown on SrTiO$ _{3} $ (001) substrate, using pulsed laser deposition.
The details of fabrication are described in Ref. \cite{Suvankar}. 
HAXPES measurements were performed at BL47XU in SPring-8 at room temperature with the photon energy of 7940 eV (probing depth is $ \sim $ 5 nm) \cite{SPring-8} for both BaFeO$ _{3} $ thin films and 50 nm thin films of SrFeO$ _{3} $ on (LaAlO$ _{3} $)$ _{0.3} $-(SrAl$ _{0.5} $Ta$ _{0.5} $O$ _{3} $)$ _{0.7} $ (LSAT) (001) substrates to compare their electronic states \cite{SuvankarSFO}.
The Fermi energy ($ E_{F} $) position and the energy resolution of $\sim$ 250 meV were determined by measuring the Fermi edge of gold which has electrical contact with the samples. 
XAS and XMCD were carried out at BL-16A in Photon Factory at a temperature of 20 K.
The magnetic field of $ \sim $ 1 T was applied perpendicular to the thin film surface, strong enough to saturate the out-of-plane magnetization \cite{Suvankar}. 
XMCD was measured by switching the applied magnetic field under a fixed circular polarization of the incident beam incident on the sample perpendicular to its plane. 
The XAS and XMCD spectra were detected with the total-electron-yield mode with the probing depth of $ \sim $ 2 nm \cite{Correction TEY, ProbingDepth}. 

The CI cluster-model calculation has been prevalently utilized to describe the physical properties of correlated materials since it can treat the electronic correlation in a cluster precisely \cite{Imada,Bocquet,Tanaka,Wadati}. 
The details of the CI cluster-model calculations are given in Refs. \cite{Bocquet,Tanaka,Wadati}. 
In this model, multiplet states including the charge transfer from oxygen ions to the iron ion are considered for a single octahedral [FeO$ _{6} $]$ ^{8-} $ cluster.
The wave function for the ground state ($ \Psi_{i} $, Eq. (1)) and for the final state of Fe $ 2p $ core-level HAXPES ($ \Psi_{f_{1}} $, Eq. (2)),   of valence-band HAXPES ($ \Psi_{f_{2}} $, Eq. (3)) and of Fe $ 2p $ XAS ($ \Psi_{f_{3}} $, Eq. (4)) are described as
	\begin{eqnarray}
	\mid{\Psi_{i}\rangle}& = &\alpha_{1}\mid{d^{4}}\rangle + \alpha_{2}\mid{d^{5}\underline{L}\rangle} + \alpha_{3}\mid{d^{6}\underline{L}^{2}\rangle} + \cdots\\ 
	\mid{\Psi_{f_{1}}\rangle}& = &\beta_{1}\mid{\underline{c}d^{4}}\rangle + \beta_{2}\mid{\underline{c}d^{5}\underline{L}\rangle} + 
	\beta_{3}\mid{\underline{c}d^{6}\underline{L}^{2}\rangle} + \cdots\\ 
    \mid{\Psi_{f_{2}}\rangle}& = &\gamma_{1}\mid{d^{3}}\rangle + \gamma_{2}\mid{d^{4}\underline{L}\rangle} + \gamma_{3}\mid{d^{5}\underline{L}^{2}\rangle} + \cdots\\
	\mid{\Psi_{f_{3}}\rangle} &=& \delta_{1}\mid{\underline{c}d^{5}}\rangle + \delta_{2}\mid{\underline{c}d^{6}\underline{L}\rangle} + 	\delta_{3}\mid{\underline{c}d^{7}\underline{L}^{2}\rangle} + \cdots
    \end{eqnarray}
respectively.
Here, $ \underline{c} $ denotes the Fe 2$ p $ core hole.
Fitting parameters in this model are the charge transfer energy from the O $ 2p $ orbitals to the empty Fe $ 3d $ orbitals denoted by $ \Delta $, the strength of Fe $ 3d $ - O $ 2p $ hybridization denoted by Slater-Koster parameters $ (pd\sigma) $, the  on-site $ 3d$ - $3d$ Coulomb interaction energy denoted by $ U_{dd} $.
The parameters used in this work are summarized in TABLE \ref{tab1} which were adjusted near the phase boundary of A-type helicoidal magnetic and ferromagnetic phases to explain the magnetic ground state of BaFeO$ _{3} $ thin films \cite{Mostovoy}.
The CI cluster-model calculation with our set of parameters revealed that the ground state is dominated by $ d^{5}\underline{L} $ configuration (17 \% $ d^{4} $, 65 \% $d^{5}\underline{L} $, 18 \% $ d^{6}\underline{L}^{2} $) due to the negative charge transfer energy.


%
\begin{figure}
\begin{center}
\includegraphics[width=7.5cm]{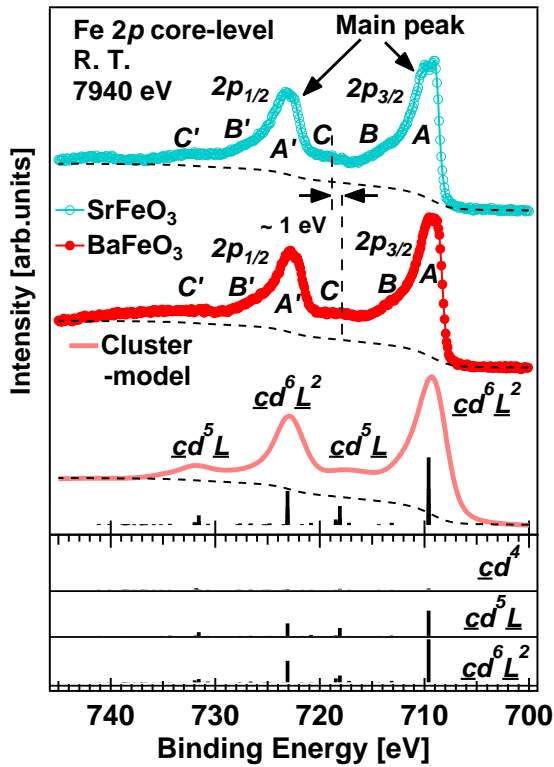}
  \caption{(Color online) Fe 2$ \textit{p} $ core-level HAXPES spectra of BaFeO$_{3}$ thin film grown on SrTiO$_{3}$ substrate (middle) compared with that of SrFeO$_{3}$ thin film on LSAT substrate (top) and CI cluster-model calculation (bottom).  The vertical bars under the bottom spectrum indicates the calculated states obtained by the CI cluster-model calculations. The states are decomposed into $ \underline{c}d^{4}$, $ \underline{c}d^{5}\underline{L}  $ and $ \underline{c}d^{6}\underline{L}^{2}  $ configurations as shown below the spectra. The dotted lines denote the background due to secondary electrons.}
  \label{BFOcore}
 \end{center}
\end{figure} 
%
%
%
\section{Results and discussion}
\begin{table}[t]
		\begin{center}
		\caption{Parameters used for the CI Cluster-model calculations of 
		HAXPES in Fig. \!\!\ref{BFOcore} and Fig. \!\!\ref{BFOvalence}, and XAS and XMCD spectra in Fig. \!\!\ref{FeXMCD}. 
	   See text for details.}
				\begin{tabular*}{8.5cm}{@{\extracolsep{\fill}}cccccl}
				\hline \hline
				  &$\Delta$ & ($pd\sigma)$ & $U_{dd}$ &  [eV]\\
				\hline
				BaFeO$_{3}$ & $ -0.9 $ & $ -1.5 $ & 7.1 &\\
				\hline \hline
				\end{tabular*}
		\label{tab1}
		\end{center}
\end{table}

Figure \ref{BFOcore} shows the Fe $2p$ core-level HAXPES spectra for SrFeO$ _{3} $ (top), BaFeO$ _{3} $ (middle) and CI cluster-model calculation (bottom).
The experimental spectra have main peaks (feature A and A$ ^{\prime} $) and weak satellite structures (feature B, C, B$ ^{\prime} $ and C$ ^{\prime} $) for both the $2p_{3/2}$ and $2p_{1/2}$ transition as indicated in the top and middle spectra of Fig. \!\!\ref{BFOcore}.
Since Fe$ ^{3+} $ would greatly enhance the feature C, the weak intensities of the feature C in our results suggest that the formal valence of Fe is not 3+ but 4+ and that both the sample SrFeO$ _{3} $ and BaFeO$ _{3} $ thin films were fully oxidized  (probing depth; $ \sim $ 5 nm from the surface) \cite{Bocquet,Suvankar,Wadati}.
The CI cluster-model calculation indicates that the main peaks (feature A and A$ ^{\prime} $) are mainly composed of the $  \underline{c}d ^{6} \underline{L}^{2} $ configuration.
The satellites of feature C and C$ ^{\prime} $ are mainly composed of $   \underline{c}d ^{5}\underline{L} $, as shown by the assignments in the bottom of Fig. \!\!\ref{BFOcore}.
The weak satellites of feature B and B$ ^{\prime} $ in the both of the experimental spectra are not reproduced by CI cluster-model calculations. 
The origins of the feature B and B$ ^{\prime} $ are most likely attributed to non-local screening effect, which is often seen in the core-level spectra of high-covalent compounds \cite{VeenendaalNi,BocquetNi}.  
The structures of the Fe $2p$ core-level of BaFeO$ _{3} $ are quite similar to that of SrFeO$ _{3} $.
The slight lowering in the binding energy of feature C in BaFeO$ _{3} $ compared to that in SrFeO$_{3}  $ originates from the reduction of Fe 3$ d $ - O $ 2p $ hybridization in BaFeO$ _{3} $ due to the lattice expansion in BaFeO$ _{3} $ following Harrison's rule for $ (pd\sigma)  \propto  R^{-3.5} $ \cite{Harrison}.

\begin{figure}[t]
 \begin{center}
  \includegraphics[width=7.5cm]{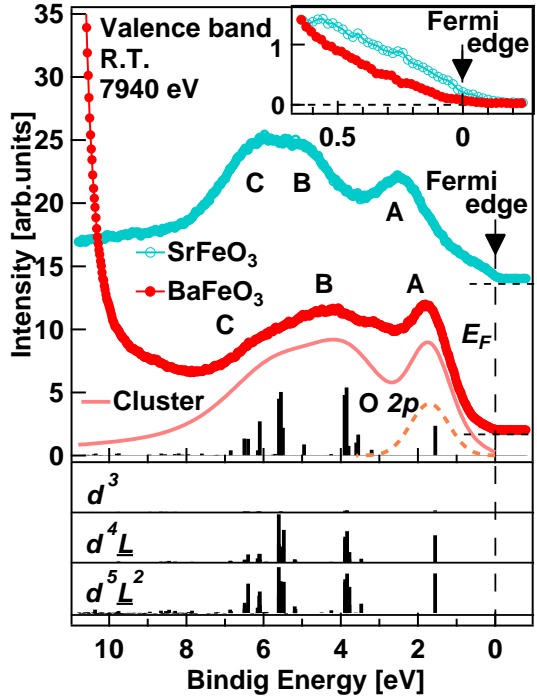}
  \caption{(Color online) HAXPES valence-band spectra of SrFeO$_{3}$ (top) and  BaFeO$_{3}$ (middle) thin films. 
  The bottom spectrum is the result of the cluster-model calculations, which the multiplet states, shown as bars below the spectrum, were broadened.
  The inset shows the expanded spectra  around $ E_{F} $. 
  The multiplet states decomposed into $ d^{3}$, $ d^{4}\underline{L}  $ and $ d^{5}\underline{L}^{2}  $ components are also shown below those spectra.}
  \label{BFOvalence}
 \end{center}
\end{figure}

Figure \ref{BFOvalence} shows the valence-band  HAXPES spectra of SrFeO$_{3}$ (top) and BaFeO$_{3}$ (middle) thin films.
We also showed the calculated multiplet state for BaFeO$ _{3} $ thin films in the bottom bars. 
The bottom bars were broadened by a Lorentzian to consider the lifetime broadening of the photohole, and by a Gaussian for the combined effects of the instrumental resolution and the $ d $ band dispersion \cite{Wadati}. 
The broadened spectrum is also shown in the bottom, which nicely reproduces the experimental result of BaFeO$ _{3} $ thin films. 
Note that we added O 2$p$ band, which is assumed to be a Gaussian, to the Fe 3$d$ spectrum to calculate the photoemission spectrum. 
The energy position of O 2$p$ band is deduced from the result of local-spin-density-approximation calculations in Ref. \cite{TheoreticalBFO2}.

 The spectra have three main structures indicated by A, B and C, respectively, besides the Fermi edge in SrFeO$ _{3} $.
A large upturn structure, at $  \sim $ 10 eV in BaFeO$_{3}$, is the tail of  Ba 5 $p_{3/2}$ core-level peak at $ \sim $ 11 eV.
The inset shows the expanded spectra near $ E_{F} $ to show the Fermi edge in SrFeO$ _{3} $ and does not in BaFeO$ _{3} $.
They clearly contrasts  insulating BaFeO$ _{3} $ thin films with metallic SrFeO$ _{3} $.
These results agree qualitatively with the resistivity measurements \cite{Suvankar,SuvankarSFO}.
Since SrFeO$ _{3} $ has stronger Fe 3$ d $ - O 2$ p $ hybridization due to the smaller lattice constant, the structure of feature A shifts  toward higher binding energy, and its bandwidth becomes broader, as we will discuss later.
These features are reproduced by local-spin-density-approximation calculations in Ref. \cite{TheoreticalBFO2}, although they predicted that BaFeO$ _{3} $ becomes half-metallic.
The multiplet states calculated by cluster-model calculations were decomposed into $d ^{3}$, $ d^{4}\underline{L}  $ and $ d^{5}\underline{L}^{2}  $ components. 
The feature B and C in the experimental spectrum for BaFeO$ _{3} $ consist of strongly hybridized states between $ d^{4}\underline{L}  $ and $ d^{5}\underline{L}^{2}  $ configurations.
The discrete state composing the feature A, which locates the binding energy of $\sim$ 1.8 eV, is created by  the strong Fe $ 3d $ - O $ 2p $ hybridization.
The results of cluster-model calculations showed that the $ d^{5}\underline{L}^{2}  $ configuration dominates at the discrete first ionization state (2 \% $ d^{3} $, 40 \% $ d^{4}\underline{L}$,  58 \% $ d^{5}\underline{L}^{2} $), indicating the heavy O 2$p$ character at the state. 
Therefore, the dominant transition from the ground state to first ionization state is $ d^{5}\underline{L}  \longrightarrow d^{5}\underline{L}^{2}  $, indicating the extra oxygen hole in this transition.

%

\begin{figure}[t]
	 \begin{center}

	   		   \includegraphics[width=7cm]{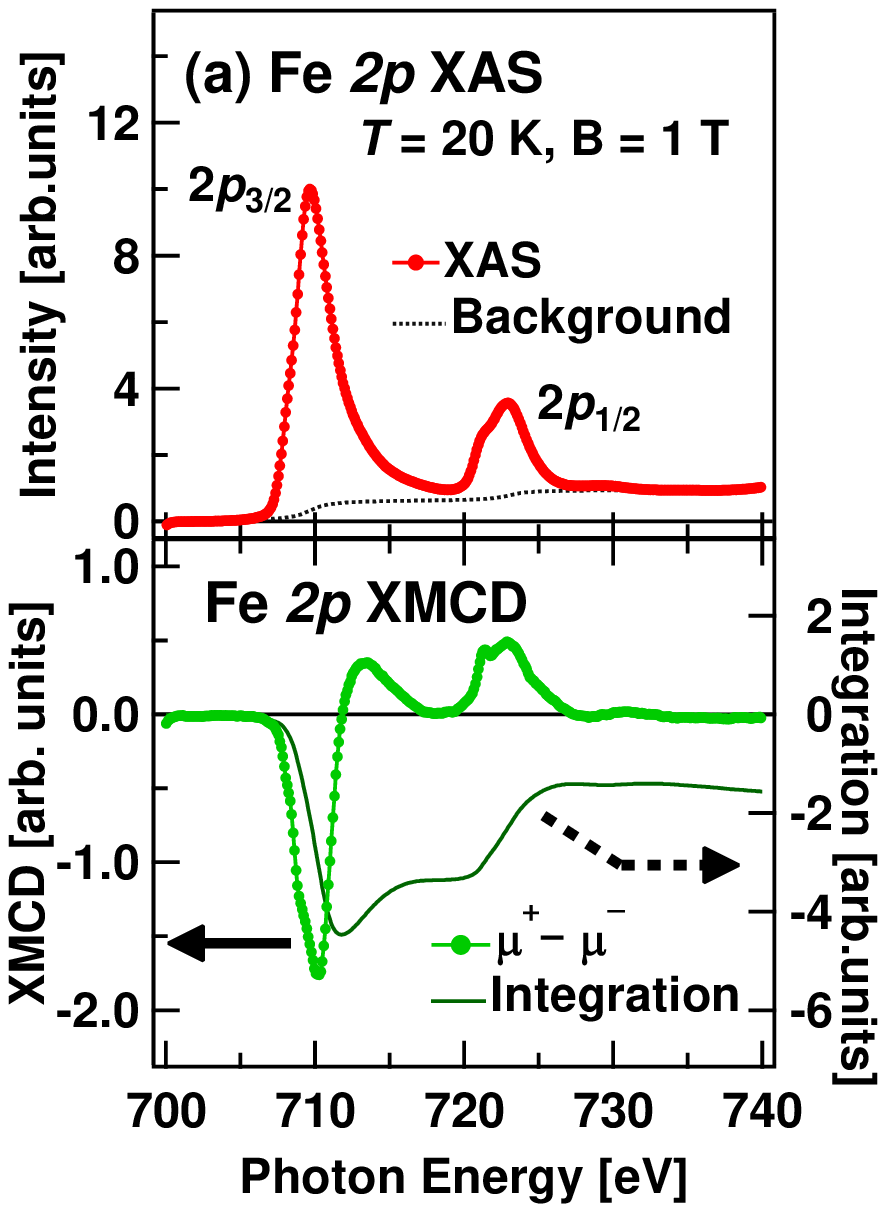}


	      \includegraphics[width=7cm]{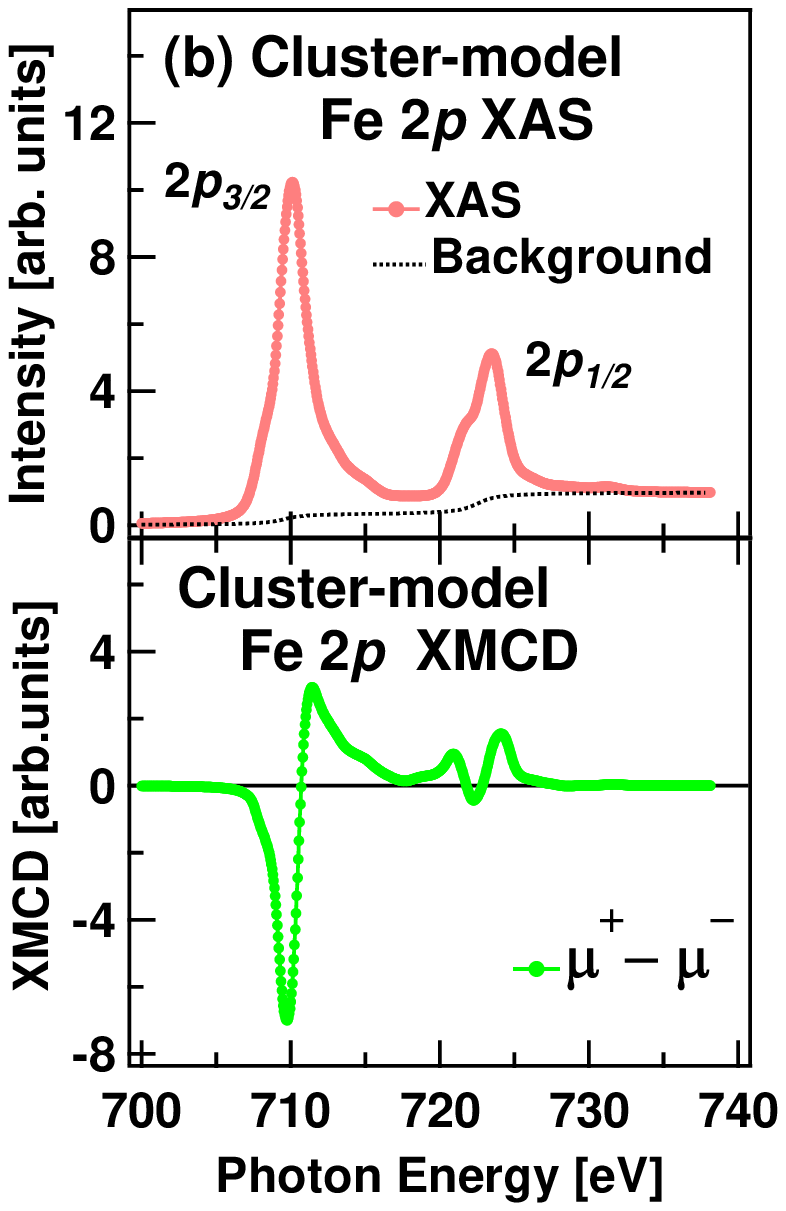}
	 		\caption{(Color online) Fe $2p$ XAS and XMCD spectra of BaFeO$_{3}$ thin films are shown in panels (a) and (b) obtained by experiment and CI cluster-model calculation, respectively. }
   	   \label{FeXMCD}
   	   \end{center} 
 \end{figure}

The Fe $2p$ XAS (top) and XMCD (bottom) spectra are shown in Fig. \!\!\ref{FeXMCD} (a). 
The broad $ 2p_{3/2} $ XAS peak without shoulder structures reflects the heavily mixed state due to the strong hybridization in BaFeO$ _{3} $ which is quite similar to that in SrFeO$ _{3} $ \cite{Abbate}. 
These broad peaks seen in BaFeO$ _{3} $ and SrFeO$ _{3} $ suggests that the valence of Fe is not 3+ but 4+ \cite{Abbate}.
Fe $2p$ XMCD shows a large peak intensity of $ \sim $ 18 \% of the XAS peak intensity. 
The XMCD intensity of BaFeO$ _{3} $ thin film is approximately twice as large as that of bulk BaFeO$ _{3} $ \cite{mizumaki}. 
To obtain the orbital moment ($ M_{orb} $) and spin moment ($ M_{spin} $), we applied XMCD sum rules in Eq. (5) and (6) \cite{SumRuleO,SumRuleS}.
	\begin{eqnarray}
	M_{orb}&=&\dfrac{2n_{h}}{3N}\int_{2p_{1\!/2}+2p_{3\!/2}}\!\!\!\!\!\!\!\!\!\!\varDelta\mu \:\:dE\\
	M_{spin}\!-\!\dfrac{7}{2}M_{T}\!\!&=&\!\dfrac{n_{h}}{N}\Bigl(\int_{2p_{3\!/2}}\!\!\!\!\!\!\!\!\!\varDelta\mu \:dE\:\: -2\!\!\int_{2p_{1\!/2}}\!\!\!\!\!\!\!\!\!\varDelta\mu\:dE\Bigr)
	\end{eqnarray}
	$\varDelta\mu = \mu^{+}-\mu^{-} $ and $\dfrac{1}{2}(\mu^{+}+\mu^{-})$ corresponds to the XMCD and XAS intensity defining the expression for $ N = \dfrac{1}{2} \int_{2p_{1\!/2}+2p_{3\!/2}}\!(\mu^{+}+\mu^{-})\:dE$. Here, $\mu^{+}$ and $\mu^{-}$ denote the absorption intensities for left-handed and right-handed circularly polarized incident photons.
$n_{h}$ denotes the number of holes in the $ d $ orbital, and we assumed $ n_{h} = 6 $ for this analysis due to the formal valence of Fe$ ^{4+} $ \cite{mizumaki}.
$M_{T} $ denotes the $z $ component of the magnetic dipole moment, which is negligible in this case because of the O$ _{h} $ crystal symmetry of BaFeO$ _{3} $ \cite{SumRuleS, TzNeglegible,J.Stohr}.

By applying the XMCD sum rules, we obtained a large magnetic moment of $M_{total}$ = $2.1\,\pm\, 0.3 \,\mu_{B}/f.u.$, composed of $M_{spin}=1.8\,\pm\, 0.2 \,\mu_{B}/f.u.$ and $ M_{orb}\,=\,0.3\,\pm\, 0.1 \,\mu_{B}/f.u.$.
Note that we considered the correction coefficient of 1/0.58 for $M_{spin}$ due to strong electronic correlations \cite{Correction}.
We also considered corrections of 1/0.88 for $M_{spin}$ and 1/0.49 for $M_{orb}$ to compensate for the saturation effect in total-electron-yield mode \cite{Correction TEY}.
Here, we assumed that the x-ray attenuation length at the $ 2p_{3/2} $ is $ \sim $ 2 nm \cite{Correction TEY, ProbingDepth}.
Since both the $M_{spin}$ ($ > $ 0) and $M_{orb}$ ($ > $ 0) have the same sign in BaFeO$ _{3} $, it suggests the substantial weight of the $d^{6}\underline{L}^{2}$ configuration in the ground state of BaFeO$ _{3} $, by taking Hund's rule into account.
Since BaFeO$ _{3} $ thin films have negative charge transfer energy, it is deduced that the $d^{6}\underline{L}^{2}$ configuration is stabilized some extent besides the main component of the $d^{5}\underline{L}$ configuration, which is consistent with the CI cluster-model calculations. The magnetic moment obtained by the experimental XMCD and superconducting quantum interference device (SQUID) measurements are shown in TABLE \ref{tab2}. 
The XMCD result gave a smaller total magnetic moment than the value obtained by SQUID measurement.
The origin of the deviation is not clear at this moment, but there is a possibility that the magnetic moment on the oxygen site might cause the underestimation based on the total magnetic moment in the iron site.

In Fig. \!\!\ref{FeXMCD} (b), the calculated Fe 2$ p $ XAS (top) and XMCD (bottom) spectra are presented.
The cluster-model spectrum reproduces the features of the experimental results very well, as oppose to the atomic multiplet calculation for $ d^{4} $ in Ref. \cite{Abbate}, suggesting the importance of charge transfer effects in this system.

\begin{figure}[t]
	\begin{center}
	 		  \includegraphics[width=7cm]{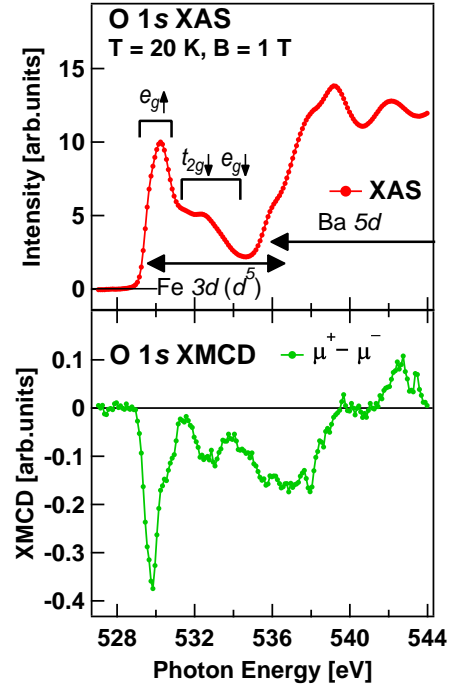}
	 		  \caption{(Color online) O $1s$ XAS and XMCD spectra of BaFeO$_{3}$ thin films are presented.}
	 		     	   \label{OXMCD}
	\end{center}
\end{figure}

The O $1s$ XAS (top) and XMCD (bottom) spectra are presented in Fig. \!\!\ref{OXMCD}.
O $ 1s $ XAS spectra of correlated compounds originate from transitions into unoccupied states with O $ 2p $ character hybridized with transition metal $ 3d $ states.
Therefore, the structure of the spectrum is qualitatively related to empty bands of primarily Fe 3$ d $ weight.
The structure from 528 eV to 534 eV, mainly Fe $ 3d $-related, is similar to that  of SrFeO$ _{3} $ \cite{Abbate}. 
The O $1s$ XAS of La$ _{1-x} $Sr$ _{x} $FeO$ _{3} $ shows that a new structure below the threshold of LaFeO$ _{3} $ grows rapidly with increasing Sr, and dominates the spectra at high concentrations, corresponding to the Fe related main structure in our spectrum \cite{Abbate}.
Those main features seen in SrFeO$ _{3} $ and BaFeO$ _{3} $ are attributed strong hybridization between Fe 3$ d $ - O 2$ p $ with e$ _{g\uparrow} $ symmetry.
Such effective states are also seen in La$ _{2-x} $Sr$ _{x} $CuO$ _{4} $ and Li$ _{x} $Ni$ _{1-x} $O due to the anomalously strong antiferromagnetic coupling between Cu or Ni $ 3d $ and O $ 2p $ \cite{LSCO,LNO}.

The O $ 1s $ XMCD spectrum directly shows that the effective state in BaFeO$ _{3} $ has O $2p$ character with the peak intensity of $ \sim $ 4 $ \% $ versus that of XAS.  
This is the evidence of the significant O $2p$ hole character in the ground states due to the negative charge transfer energy and strong Fe $ 3d $ - O $ 2p $ hybridization.
Since the threshold of XAS is dominated by the transition of $ d^{5}\underline{L}  \longrightarrow  \underline{1s}d^{5} $, the strong intensity of XMCD was observed at the XAS structure of the lowest energy even in the usually nonmagnetic oxygen.
Here, $ \underline{1s} $ denotes the O 1$ s $ core hole created by the incident x-ray. 
Thus, when an extra electron is added to the ground state, the electron will be introduced into the ligand hole at the affinity level.


\begin{table}[t]
	\begin{flushleft}
		\caption{Magnetization obtained by XMCD and SQUID. We applied the XMCD sum rules to obtain spin, magnetic and total magnetic moments from Fe $2p$ XMCD. The SQUID measurement is taken from Ref. \cite{Suvankar}.}
				\begin{tabular*}{8.5cm}{@{\extracolsep{\fill}}lcccl}
				\hline\hline
				 & M$_{spin}$ & M$_{orb}$ & M$_{total}$&[$\mu_{B}/f.u.]$\\
				\hline
				XMCD& $1.8 \pm 0.2$ & $0.3 \pm 0.1$ & $2.1 \pm 0.3$ &\\
                SQUID           &\textemdash&       \textemdash      & $2.7$  &       \\
				\hline\hline
				\end{tabular*}
		\label{tab2}
	\end{flushleft}
\end{table}

\begin{figure}
 \begin{center}
  \includegraphics[width=9cm]{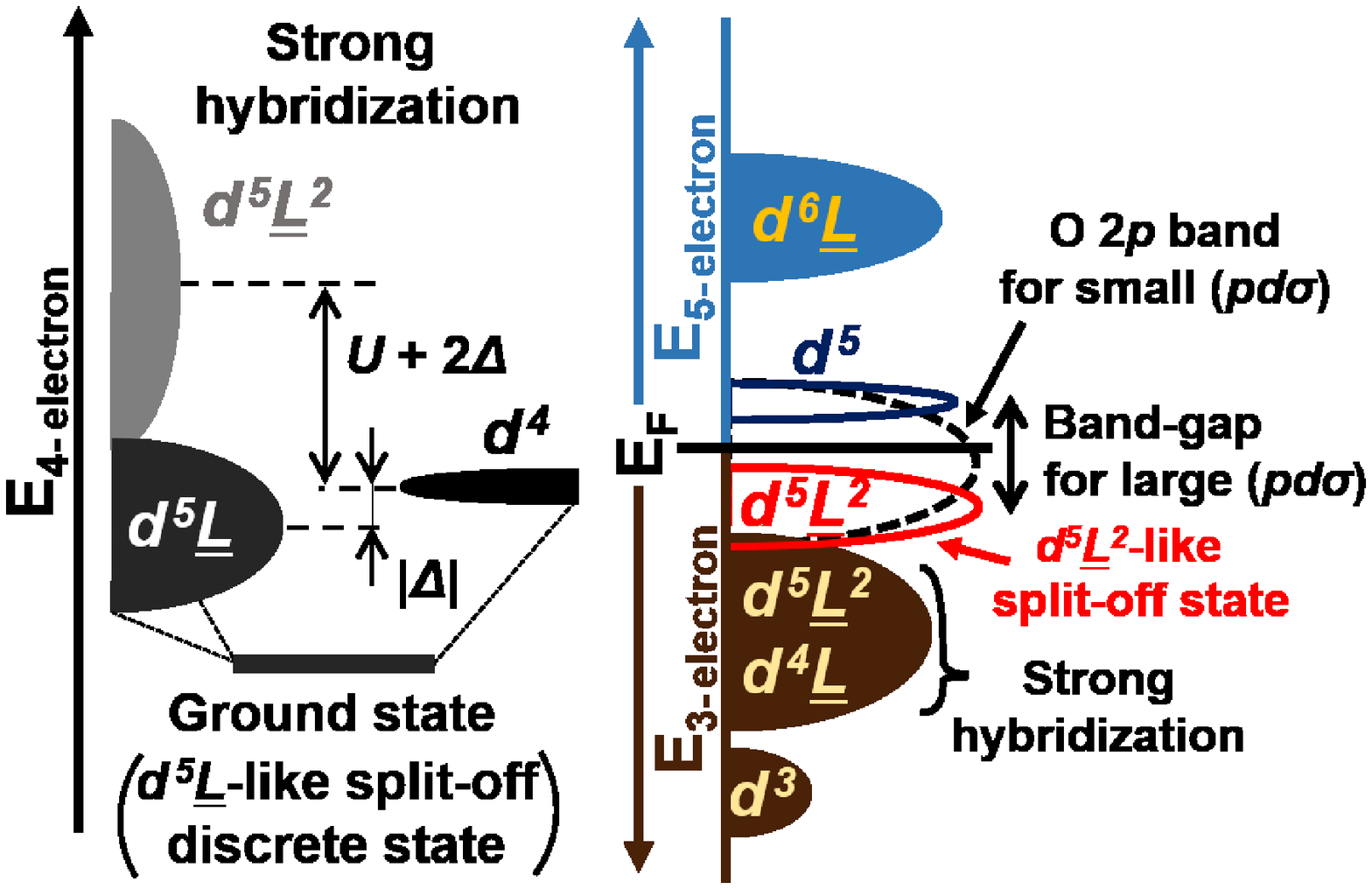}
  \caption{(Color online) Energy levels for the 4-electron system (left) and the density of states for BaFeO$ _{3} $ are schematically shown (right panel). Due to the strong Fe 3$ d $ - O 2$ p $ hybridization, the $ d^{5}\underline{L} $-like split-off state is created under the continuum $ d^{5}\underline{L} $ state (left panel). As a result, the band gap of $d^{5} $ + $d^{5}\underline{L}^{2}$ is formed (right panel) due to the significant stabilization of the discrete state (left panel).}
  \label{GAP}
 \end{center}
\end{figure}
We showed that the experimental spectra of BaFeO$ _{3} $ are reproduced quite well by the results of CI cluster-model calculations with a common set of parameters, which can explain the magnetic ground state of BaFeO$ _{3} $ by comparing to the phase diagram in Ref. \cite{Mostovoy}. 
	The CI cluster-model calculations clarified that the ground state, the first affinity level and first ionization level are dominated by $ d^{5}\underline{L} $ (17 \% $ d^{4} $,  65 \% $ d^{5}\underline{L}$, 18 \% $ d^{6}\underline{L}^{2}$), by $ d^{5} $ (71 \% $ d^{5} $, 26 \% $ d^{6}\underline{L} $, 3 \% $ d^{7}\underline{L}^{2} $) and by $ d^{5}\underline{L}^{2} $ (2 \% $ d^{3}$, 40 \% $ d^{4}\underline{L} $, 58 \% $ d^{5}\underline{L}^{2} $).
	Therefore, the band gap is characterized by $ d^{5}\underline{L} $ + $ d^{5}\underline{L}  \longrightarrow  d^{5} $ + $d^{5}\underline{L}^{2} $ and has significant O 2$ p $ character, which is schematically shown in the right panel of Fig. \!\!\ref{GAP} \cite{wadati2}.
From those results and the set of the parameters for the CI cluster-model calculation, BaFeO$ _{3} $ is classified as a  covalent insulator (negative-charge-transfer-energy insulator) as proposed by Sarma \textit{et al}.  (Mizokawa \textit{et al}.) in the modified Zaanen-Sawatzky-Allen diagram. \cite{Imada,SarmaCI,SarmaMZSA,SarmaMZSA2, mizokawa,mizokawa2}. 
Zaanen-Sawatzky-Allen diagram, which classifies the electronic properties of correlated materials systematically, was modified to consider the insulating phase due to the strong covalency within the negative-charge-transfer-energy region \cite{SarmaMZSA,SarmaMZSA2,mizokawa2}.

The $ d^{5}\underline{L} $ configuration composing the ground state is originally a continuum and thus it tends to be metallic. 
The strong Fe $ 3d $ - O $ 2p $ hybridization, however, creates the $ d^{5}\underline{L} $-like split-off state below the continuum $ d^{5}\underline{L} $ structure with significant stabilization, as schematically shown in the left panel in Fig. \!\!\ref{GAP} \cite{mizokawa2}. 
As a result, the oxygen hole is strongly localized despite the weak electronic correlation on the oxygen site.

Due to the localized oxygen hole, the system becomes an insulator and the band gap opens, as shown in the right panel of Fig. \!\!\ref{GAP} \cite{mizokawa2}.
NaCuO$ _{2} $ (Cu$ ^{3+} $) has also localized oxygen holes due to anomalously strong antiferromagnetic Cu 3$ d $ - O 2$ p $ hole coupling, whose split-off $ d^{9}\underline{L} $-like state with the $ ^{1} A_{1}$ symmetry is denoted as the Zhang-Rice singlet \cite{mizokawa}.

Although the discrete split-off state is stable within a single cluster, the band gap tends to become unstable and collapse in a periodic lattice, unless the interactions between the clusters are weak, such as for the distorted bond angle structure in CaFeO$ _{3} $, or orthogonality of the orbitals between the clusters in NaCuO$ _{2} $ \cite{mizokawa2}. 
However, the small band gap in CaFeO$ _{3} $  by  small distortions, unlike the strong one in NaCuO$ _{2} $, induces charge disproportionation by overcoming the band gap of  $d^{5}\underline{L}\,+\,d^{5}\underline{L}$ $\longrightarrow$ $d^{5}$ + $d^{5}\underline{L}^{2}$ \cite{CFO00,mizokawa2}.
On the other hand, the non-distorted structure in SrFeO$ _{3} $, with the Fe-O-Fe bond angle of $ \sim\,180^{\circ}$, completely closes the band gap to form a metal in a periodic lattice. 
According to the modified Zaanen-Sawatzky-Allen diagram, the boundary of the metal-insulator transition is sensitive not only to $\Delta$ and $U$, but also ($pd\sigma$) and O $ 2p $ bandwidth denoted by W$ _{p} $, within the negative-charge-transfer-energy regime \cite{SarmaMZSA,SarmaMZSA2,mizokawa2}. 
In this regime, the band gap tends to be larger as ($ pd\sigma $) increases and as W$ _{p} $ decreases \cite{mizokawa2}.

In the case of BaFeO$ _{3} $ and SrFeO$ _{3} $, the metal-insulator transition occurs due to the competition of the parameters of ($ pd\sigma $) and W$ _{p} $.
We should note that the effective one-electron bandwidth is approximately related to ($ pd\sigma $)  by W$ _{p} $ $ \propto $ $ (pd\sigma)^{2} $ via the second-order process. 
Since the ratio of atomic distances in BaFeO$ _{3} $ to SrFeO$ _{3} $ is $\left( \dfrac{R_{BFO}}{R_{SFO}}\right) $  $ \sim $ 1.03, we obtained the rough estimation of the ratio of $ \dfrac{(pd\sigma)_{BFO}}{(pd\sigma)_{SFO}}\,\sim\,\left( \dfrac{R_{BFO}}{R_{SFO}}\right) ^{-3.5}$ $ \sim $ 0.91 and of $ \dfrac{\left( W_{p}\right)_{BFO}}{\left(W_{p}\right)_{SFO}}\,\sim\,\left( \dfrac{R_{BFO}}{R_{SFO}}\right) ^{-7}$ $ \sim $ 0.82, from Harrison's rule \cite{Harrison}.
The band gap in BaFeO$ _{3} $ is explained by the significant reduction of the bandwidth compared to the relatively weaker reduction of the Fe $3d$ - O $2p$ hybridization. 
We can see directly the differences of those parameters in their valence-band spectra in Fig.\!\! \ref{BFOvalence}. 
The feature A of SrFeO$ _{3} $ is located at a higher binding-energy position than that of BaFeO$ _{3} $. 
Since the Fe 3$ d $ - O 2$ p $ hybridization is stronger in SrFeO$ _{3} $, the split-off discrete state of 4-electron systems (see left panel in Fig.\!\! \ref{GAP}) is more stable than that of BaFeO$ _{3} $. 
As a result, the energy difference of the ground state in 4-electron systems and feature A in 3-electron systems for SrFeO$ _{3} $ tends to be larger than that for BaFeO$ _{3} $. 
The broader bandwidth of feature A in SrFeO$ _{3} $ valence-band spectrum compared to that in BaFeO$ _{3} $ also suggests the stronger Fe 3$ d $ - O 2$ p $ hybridization in SrFeO$ _{3} $.
Even in the non-distorted structure in BaFeO$ _{3} $ and in a periodic lattice, the band gap remains open since the interactions between the clusters are weak due to the larger lattice constant.

To summarize, as the lattice constant increases from CaFeO$ _{3} $ to SrFeO$ _{3} $, the Fe-O-Fe bond angle straightens to $\sim$ 180$^{\circ}$ and the O $ 2p $ bandwidth increases.
Therefore, SrFeO$ _{3} $ is metallic. 
When the lattice constant is further increased from SrFeO$ _{3} $ to BaFeO$ _{3} $ the effect of narrowing bandwidth causes a further transition to an insulator.

The coefficient of antiferromagnetic superexchange interaction ($ J $) greatly reduces as the lattice constant increases. By considering Harrison's rule, we can roughly estimate the ratio of $ J $ between BaFeO$ _{3} $ and SrFeO$ _{3} $ as $ \sim $ $ \dfrac{\left( J\right)_{BFO}}{\left( J\right)_{SFO}}\,\sim\,\left( \dfrac{R_{BFO}}{R_{SFO}}\right) ^{-14}$ $ \sim $ 0.67 \cite{Harrison}. 
This reduction of $ J $ in BaFeO$ _{3} $ induces stabilization of ferromagnetic or A-type helicoidal magnetic phases, compared to SrFeO$ _{3} $ \cite{Mostovoy}. 
 Mostovoy explained that the ferromagnetic phase in Fe$ ^{4+} $ perovskite crystal structure with negative charge transfer energy is unstable, due to the high density of spin-flip excitations of holes from the spin-down Fermi-sea to pure oxygen spin-up bands \cite{Mostovoy}.  
	On the other hand, if the spin-flip excitation is suppressed by opening a band gap in the insulating phase, the ferromagnetic phase can be stable.                                                                                                      
	Thus, the reduction of antiferromagnetic superexchange interaction and the insulating phase of BaFeO$ _{3} $ together stabilize the ferromagnetic phase. 
 
%

%
%
\section{Conclusion}
We investigated the electronic and magnetic properties of BaFeO$_{3}$ thin films by HAXPES, XAS, XMCD and CI cluster-model analysis.
We employed the parameters for CI cluster-model calculations on the phase boundary of ferromagnetic and A-type helicoidal magnetic phase in order to explain the magnetic properties of BaFeO$ _{3} $.
The calculated results reproduced all of the experimental spectra quite well.
The CI cluster-model calculation suggests that BaFeO$ _{3} $ thin film has negative charge transfer energy, and its ground state is dominated by the $d^{5}\underline{L}$ electronic configuration.
The valence-band HAXPES showed the insulating property of BaFeO$ _{3} $ thin films. 
The band gap has heavy O $ 2p $ character, suggesting that BaFeO$ _{3} $ is a covalent insulator.
We obtained the saturation magnetic moment of $2.1\,\pm\, 0.3 \,\mu_{B}/f.u.$ from Fe $ 2p $  XMCD and found that it is composed of large spin magnetic moment and small orbital magnetic moment.
Such a  stable ferromagnetic phase under small magnetic field derives from the reduction of the antiferromagnetic superexchange interaction and  the suppression of the spin-flip excitation due to the band gap in BaFeO$ _{3} $ thin films.
Thus, the metal-insulator transition and magnetic phase transition between SrFeO$ _{3} $ and BaFeO$ _{3} $ derive from the difference of their lattice constants, and the resultant bandwidth tuning
%
%

%
%
We thank K. Amemiya for experimental support. We also thank A. Fujimori, T. Tohyama, Z. Li and  D. D. Sarma for informative discussions.  H. Y. H. acknowledges support by the Department of Energy, Basic Energy Sciences, Materials Science and Engineering Devision, under contract  DE-AC02-76SF00515. This research was supported by the Japan Society the Promotion of Science (JSPS) through the Funding Program for World-Learning Innovative R\&D on Science and Technology (FIRST program), JSPS Giant-in-Aid for Scientific Research  and by Grant for Basic Science Research Projects from the Sumitomo Foundation. 
The experiments with synchrotron radiation were carried out under the approval of SPring-8 Japan Synchrotoron Radiation Research Institute (Proposal Nos. 2012A1624 and 2012B1757) and Photon Factory Program Advisory Committee (Proposal Nos. 2013G058).
%
\bibliography{LVO1tex}

\end{document}